\documentclass[%
 reprint,
 amsmath,amssymb,
 aps,
]{revtex4-1}

\usepackage{graphicx}
\usepackage{dcolumn}
\usepackage{bm}

\usepackage[normalem]{ulem}

\begin{document}

\preprint{APS/123-QED}

\title{High-pressure study of the ground- and superconducting-state properties of CeAu$_2$Si$_2$}

\author{Gernot W. Scheerer}
 \email{gernot.scheerer@unige.ch}
 \affiliation{DQMP - University of Geneva, 24 Quai Ernest-Ansermet, 1211 Geneva 4, Switzerland.}

\author{Ga\'{e}tan Giriat}
 \affiliation{DQMP - University of Geneva, 24 Quai Ernest-Ansermet, 1211 Geneva 4, Switzerland.}

\author{Zhi Ren}
 \affiliation{Institute for Natural Sciences, Westlake Institute for Advanced Study, Hangzhou, P. R. China.}

\author{G\'{e}rard Lapertot}
 \affiliation{SPSMS, UMR-E 9001, CEA-INAC/UJF-Grenoble 1, 38054 Grenoble, France.}

\author{Didier Jaccard}
 \affiliation{DPMQ - University of Geneva, 24 Quai Ernest-Ansermet, 1211 Geneva 4, Switzerland.}

\date{\today}

\begin{abstract}
The pressure-temperature phase diagram of the new heavy-fermion superconductor CeAu$_2$Si$_2$ is markedly different from those studied previously.
Indeed, superconductivity emerges not on the verge but deep inside the magnetic phase, and mysteriously $T_{\mathrm{c}}$ increases with the strengthening of magnetism.
In this context, we have carried out ac calorimetry, resistivity, and thermoelectric power measurements on a CeAu$_2$Si$_2$ single crystal under high pressure.
We uncover a strong link between the enhancement of superconductivity and quantum-critical-like features in the normal-state resistivity.
Non-Fermi-liquid behavior is observed around the maximum of superconductivity and enhanced scattering rates are observed close to both the emergence and the maximum of superconductivity.
Furthermore we observe signatures of pressure- and temperature-driven modifications of the magnetic structure inside the antiferromagnetic phase.
A comparison of the features of CeAu$_2$Si$_2$ and its parent compounds CeCu$_2$Si$_2$ and CeCu$_2$Ge$_2$ plotted as function of the unit-cell volume leads us to propose that critical fluctuations of a valence crossover play a crucial role in the superconducting pairing mechanism.
Our study illustrates the complex interplay between magnetism, valence fluctuations, and superconductivity.
\end{abstract}
                       
\maketitle

\section{Introduction}

\begin{figure}[t]
\centering
\includegraphics[width=0.85\linewidth]{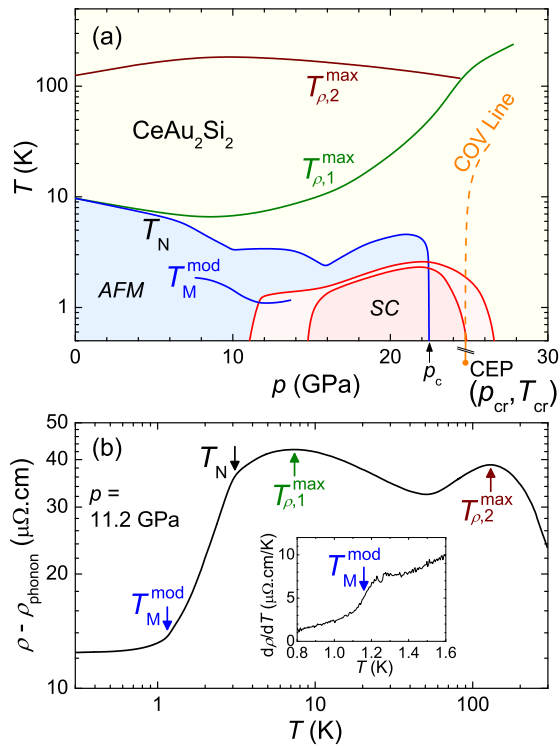}
\caption{
(a) Schematic pressure-temperature phase diagram of CeAu$_2$Si$_2$ based on Refs. \cite{ren14} and \cite{ren15} and this work.
(b) Resistivity $\rho$ versus temperature $T$ of CeAu$_2$Si$_2$ at 11.2~GPa. The phonon term was subtracted. Inset: d$\rho(T)$/d$T$ around the anomaly at $T_{\mathrm{M}}^{\mathrm{mod}}$.
}
\label{intro}
\end{figure}

Although known about for many years, heavy fermion (HF) systems are still intensely studied since they unite some of the most interesting features of strongly correlated electron systems such as quantum criticality, non-Fermi-liquid behavior, and unconventional superconductivity.
In this context, the antiferromagnet CeAu$_2$Si$_2$ presents new challenges to the current understanding of HF superconductivity.
In CeAu$_2$Si$_2$ \cite{ren14}, the magnetic and superconducting phases overlap across an unprecedentedly broad pressure range from $p=11.8$~GPa up to the critical pressure $p_{\mathrm{c}}\sim 22.5$~GPa, where the superconducting transition temperature $T_{\mathrm{c}}$ is highest, as shown in the schematic phase diagram in Fig.~\ref{intro}(a).
Contrary to all previous observations, both $T_{\mathrm{c}}$ and the magnetic ordering temperature increase simultaneously over an extended pressure range.
Thus, at low temperatures, CeAu$_2$Si$_2$ behaves very differently from the isostructural and isoelectronic parent compounds CeCu$_2$Si$_2$ and CeCu$_2$Ge$_2$. In contrast, all three systems exhibit remarkably similar properties at intermediate and high temperatures.

Among the Ce-based HF superconductors, CeAu$_2$Si$_2$ displays one of the highest $T_{\mathrm{c}}$ of $\approx2.5$~K. Superconductivity (SC) appears to involve HF quasiparticles with an effective mass $m$*$\approx100$~$m_{\mathrm{e}}$, as inferred from the large initial slope of the upper critical field \cite{ren14}.
Furthermore, the strong pair-breaking effect of non-magnetic impurities attests unconventional SC \cite{ren15}.
CeAu$_2$Si$_2$ has a tetragonal ThCr$_2$Si$_2$-type structure with the space group I4/mmm ($D^{17}_{4\mathrm{h}}$)\cite{rossi79,grier84}
and orders antiferromagnetically below $T_{\mathrm{N}}=9.6$~K\cite{book82,ota09}.
The resistivity [Fig.~\ref{intro}(b)] and thermopower exhibit typical Kondo lattice behavior \cite{amato85,garde94,link97}.
The ambient-pressure ground state is an A-type antiferromagnetic (AF) structure with a magnetic moment oriented along the \textbf{c}-axis of ($1.29\pm0.05$)~$\mu_B$ at 5~K \cite{grier84}.
The Sommerfeld coefficient $\gamma=11$ mJ/K$^2\cdot$mol and the AF energy gap $\Delta=7.5$~K are determined by fitting the low-temperature ($T<5$~K) specific heat \cite{ota09}.
The valence of the Ce ion is 3.00 \cite{grier84} and the estimated Kondo temperature is $T_{\mathrm{K}}=1.7$~K \cite{severig89a}.
$T_{\mathrm{K}}$ increases by roughly 1-2 orders of magnitude when a hydrostatic pressure of 15 GPa is applied \cite{ren16}.
Inelastic magnetic peaks in the neutron scattering spectra show crystal field excitations \cite{grier88,severig89}.
A magnetic field of 5.5 T applied along the magnetic easy axis \textbf{c} is sufficient to suppress the AF order and leads to a polarized paramagnetic state via a metamagnetic transition \cite{fujiwara06,sefat08,ota09}.
Furthermore, as observed in specific heat and magnetization experiments \cite{ota09}, CeAu$_2$Si$_2$ exhibits two successive magnetic transitions at 8.0 and 9.6~K, which separate the AF ground state from the paramagnetic state.

Concerning the effect of pressure, it is noteworthy that the AF transition line $T_{\mathrm{N}}$ is nonmonotonic forming a triple dome [see Fig.~\ref{intro}(a)], which suggests pressure-induced modification of the magnetic structure.
Moreover, the crossover line corresponding to the delocalization of 4f electrons (labeled COV below) is located just above $p_{\mathrm{c}}$ [see Fig.~\ref{intro}(a)] \cite{ren14,ren15}. The rapid collapse of magnetism and the SC enhancement may be driven by critical valence \cite{holmes07} or even orbital \cite{ren14} fluctuations.

To refine the high-pressure phase diagram of CeAu$_2$Si$_2$, we performed measurements up to 24.3~GPa on a single crystal using techniques presented in Refs. \cite{link96,holmes04} providing nearly hydrostatic pressure conditions.
Using a multiprobe setup, we examined the in-plane electric resistivity, in-plane thermoelectric power, and ac calorimetry. With magnetic fields applied along the c-axis, we examined the pressure dependence of the superconducting upper critical field $H_{\mathrm{c2}}$ and its initial slope $d H_{\mathrm{c2}}/d T|_{T\rightarrow T_{\mathrm{c}}}$, as well as the pressure dependence of the normal-ground-state properties for $H>H_{\mathrm{c2}}$.
The present work confirms the existing phase diagram \cite{ren14} and reveals the following major novelties.
i) A new transition line $T_{\mathrm{M}}^{\mathrm{mod}}$, presumably due to the reconstruction of magnetic order, is observed inside the AF phase [see Fig.~\ref{intro}(a)].
Corresponding anomalies are clearly seen, e.g. in the resistivity [Fig.~\ref{intro}(b) and inset].
ii) Intriguingly, while $T_{\mathrm{N}}$ decreases with increasing pressure, the magnetic phase becomes much more stable in a magnetic field.
iii) Comparing the magnetic-transition lines with the large superconducting dome reveals that both the emergence of SC deep inside the AF phase and the maximum $T_{\mathrm{c}}$ occur close to magnetic instabilities in two ``critical'' pressure regions.
The first region around 14~GPa corresponds to a minimum in $T_{\mathrm{N}}$ and enhanced electronic scattering rates.
The second one around 22~GPa is marked by the abrupt vanishing of magnetism, quantum-critical-like signatures, and non-Fermi-liquid (NFL) behavior.
This corroborates the idea that the SC in CeAu$_2$Si$_2$ is driven by quantum critical fluctuations.

This paper is organized as follows. Section \ref{II} presents the experimental methods. In Sect. \ref{III}, we present the results of our multiprobe experiment.
Section \ref{IIIa} focuses on the identification of the phase transitions in the pressure region of emerging SC, Sect. \ref{IIIb} shows details of the thermopower,
and Sect. \ref{IIIc} presents the pressure-temperature phase diagram and the normal-ground-state properties. In Sect. \ref{IV}, we discuss the experimental results and the possible superconducting pairing mechanism.

\section{Experiment} \label{II}
 
We performed measurements on a Sn-flux-grown CeAu$_2$Si$_2$ single crystal in a standard dilution fridge equipped with a superconducting magnet coil with a maximum field of 8.5 T.
Details of the crystal growth are described in Ref. \cite{ren14}.
To generate pressure, we used a Bridgman-type
sintered-diamond-anvil pressure cell with a pyrophyllite gasket and steatite as a soft-solid pressure medium.
Pressure was determined from the resistive $T_{\mathrm{c}}$ of a lead strip. The pressure gradient along the sample, estimated from the Pb-transition width, slowly increased from $\approx0.5$~GPa at low pressure to $\approx0.8$~GPa at maximum pressure.
The four-point resistivity was measured with a dc current along the sample basal plane.
For dc thermoelectric power and ac calorimetry measurements, a local heater very close to one sample extremity provided a temperature gradient and temperature oscillations, respectively.
The thermoelectric voltages were measured by pairs of Au and AuFe$_{0.07\%}$ wires. Technical details can be found in Refs. \cite{ren14,link96} and \cite{holmes04}.
The superconducting transition temperatures are defined as i) $T_{\mathrm{c}}^{\mathrm{onset}}$ at twice the noise deviation from a straight tangent to the normal-state resistivity just above $T_{\mathrm{c}}$ and ii) $T_{\mathrm{c}}^{\mathrm{bulk}}$ at the midway anomaly in calorimetry.
Quantities labeled as normal-state properties were obtained by applying a magnetic field corresponding to the upper critical field of the resistive-SC-transition onset ($\approx1.4\cdot H_{\rm{c2}}$, see Fig.~\ref{mmm} later).
It was verified experimentally that the normal-state properties are qualitatively independent of the magnetic field up to 8.5~T.

\section{Results} \label{III}

\subsection{Magnetic transitions} \label{IIIa}

\begin{figure}[t] 
\centering
\includegraphics[width=0.85\linewidth]{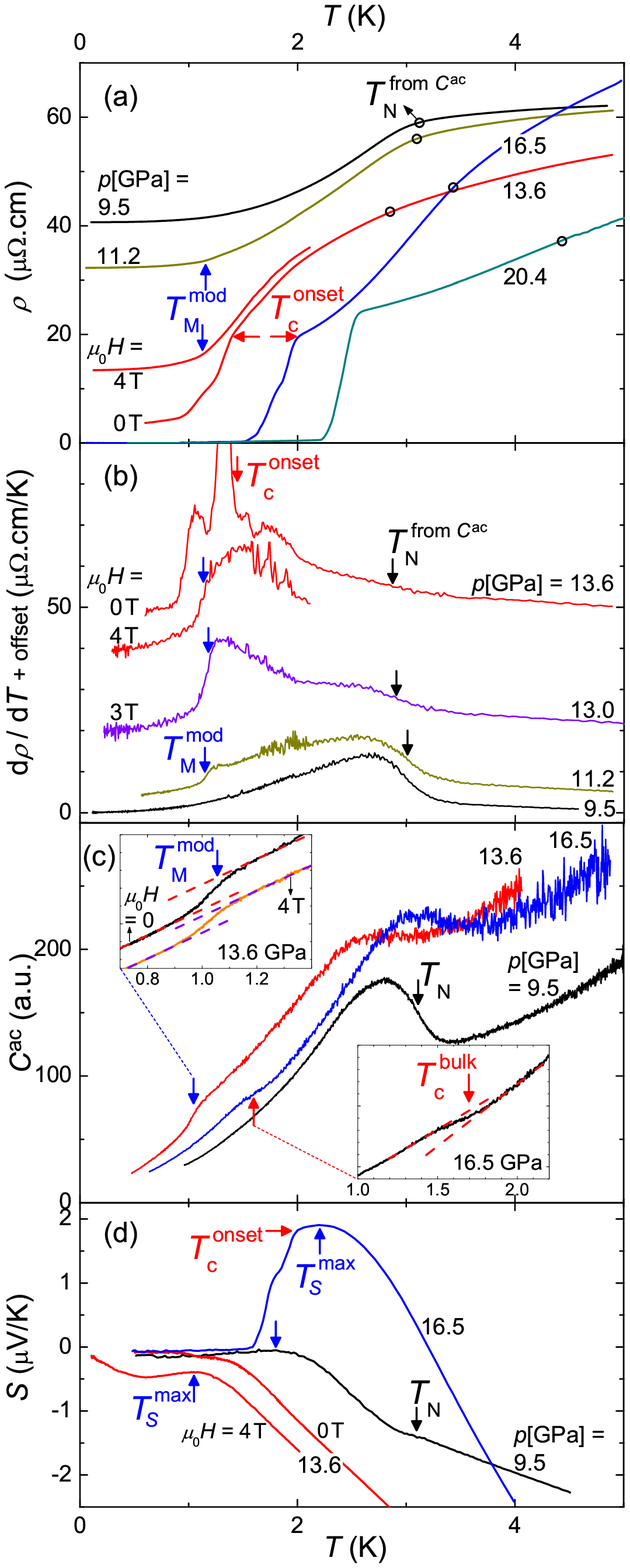}
\caption{Multiprobe data of CeAu$_2$Si$_2$ at selected pressures and at zero field (otherwise indicated).
Anomalies indicated by arrows are described in the text.
(a) In-plane resistivity $\rho$ of CeAu$_2$Si$_2$ as a function of temperature $T$. For clarity, offsets of 30 and 20~$\mu\Omega$.cm are added to the curves for 9.5 and 11.2~GPa, respectively.
The N\'{e}el temperature $T_{\mathrm{N}}$ deduced from calorimetry is indicated by circles.
(b) $d\rho/dT$ versus $T$. For clarity, offsets are added to the curves for $p\geq11.2$~GPa.
(c) Ac calorimetry $C^{\mathrm{ac}}$ versus $T$. The upper (lower) inset shows a close up of the curve at 13.6 (16.5)~GPa around the temperature $T_{\mathrm{M}}^{\mathrm{mod}}$ ($T_{\mathrm{c}}^{\mathrm{bulk}}$).
(d) Thermoelectric power $S$ versus $T$.
}
\label{data}
\end{figure}

Figure~\ref{data} displays low-temperature data measured with the multiprobe setup in the vicinity (i.e., below and above the pressure) of the SC emergence.
Generally, the anomaly at the N\'{e}el temperature $T_{\mathrm{N}}$ is clearly visible in the ac calorimetry and resistivity.
At 13.6~GPa, resistivity [Fig.~\ref{data}(a)] shows an incomplete superconducting transition with $T_{\mathrm{c}}^{\mathrm{onset}}\approx 1.4$~K,
while at 16.5~GPa, the transition is complete and also seen in the thermopower [Fig.~\ref{data}(d)].
Moreover, calorimetry [Fig.~\ref{data}(c)] has the signature of bulk SC at $T_{\mathrm{c}}^{\mathrm{bulk}}=1.6$~K.
$T_{\mathrm{c}}^{\mathrm{bulk}}$ ($T_{\mathrm{c}}^{\mathrm{onset}}$) reaches its maximum of 2.2~K (2.6~K) at 20.4~GPa.

In addition to the superconducting and AF transitions, the probes reveal other anomalies. At a temperature $T_{\mathrm{M}}^{\mathrm{mod}}$, the calorimetry exhibits a bump [see Fig.~\ref{data}(c) and upper inset] and the resistivity exhibits a kink, corresponding to an abrupt jump in the temperature derivative $d\rho/dT$ [see Fig.~\ref{data}(b)].
These anomalies are observed in the pressure range 7.6 -- 13.6~GPa, i.e., well inside the AF phase.
Their sharpness is characteristic of a phase transition and it is clear that they are not due to SC since they persist in magnetic fields higher than $H_{\rm c,2}$. 
As discussed below, the anomalies in $C^{\mathrm{ac}}$ and $d\rho/dT$ correspond to the same phenomenon, most likely to a magnetic transition (therefore the label $T_{\mathrm{M}}^{\mathrm{mod}}$).
Moreover, the thermopower [Fig.~\ref{data}(d)] shows a local maximum at $T_{S}^{\mathrm{max}}$ close to $T_{\mathrm{M}}^{\mathrm{mod}}$, which persists over the whole pressure range.

\begin{figure}[t] 
\centering
\includegraphics[width=0.85\linewidth]{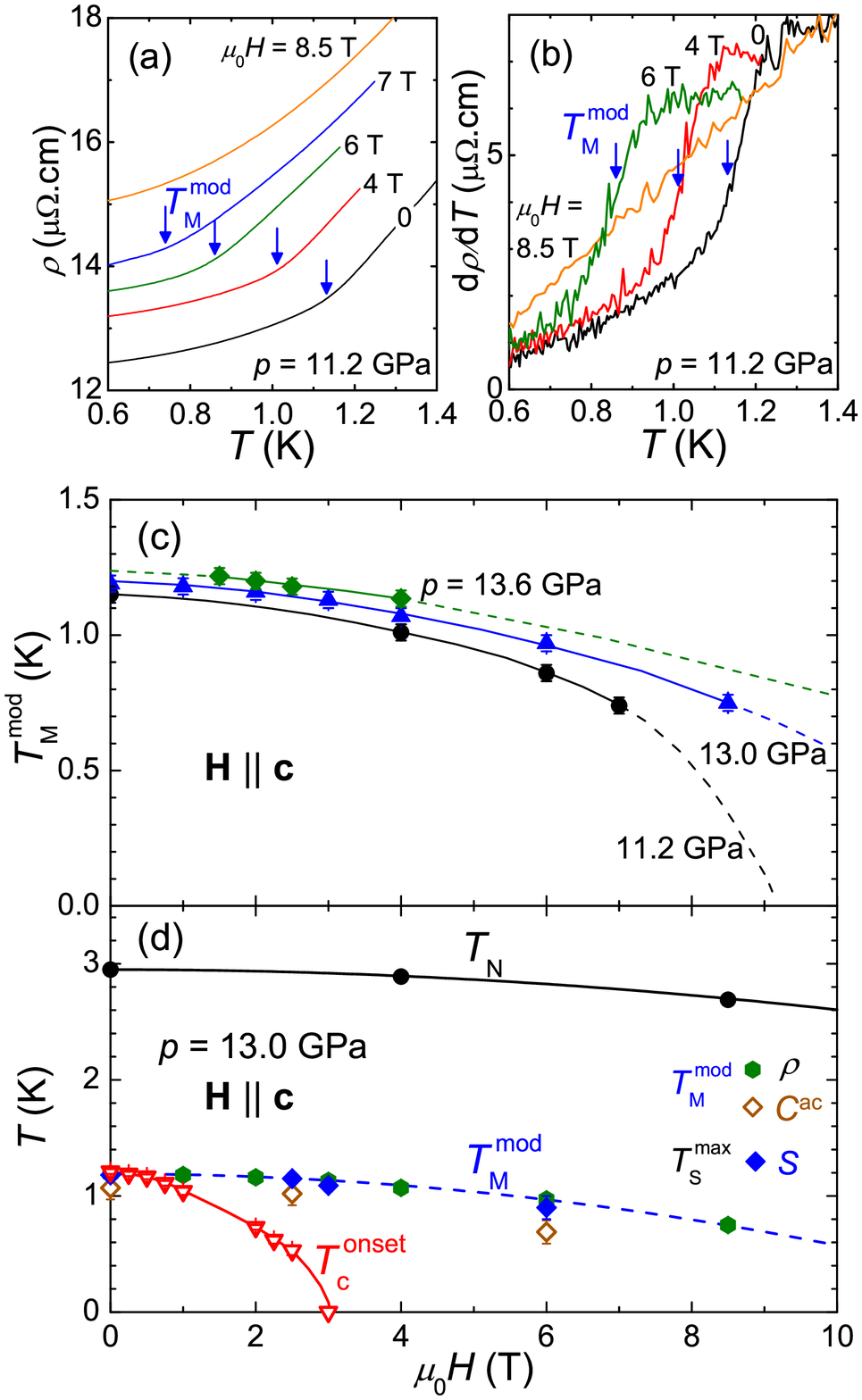}
\caption{
(a) $\rho(T)$ and (b) $d\rho/dT(T)$ at $p=11.2$~GPa at different magnetic fields $\mu_0\mathbf{H}\parallel\mathbf{c}$ up to 8.5 T. $T_{\mathrm{M}}^{\mathrm{mod}}$ is indicated by arrows.
(c) $T_{\mathrm{M}}^{\mathrm{mod}}$ deduced from $d\rho/dT$ versus $H$ applied along \textbf{c} at $p=11.2$, 13.0, and 13.6~GPa. Errors are a slightly larger than the symbol size. Lines are guides to the eyes.
(d) At $p=13.0$~GPa, field dependences of the N\'{e}el temperature $T_{\mathrm{N}}$ extracted from $C^{\mathrm{ac}}$, $T_{\mathrm{M}}^{\mathrm{mod}}$ from $C^{\mathrm{ac}}$ and $\rho$, $T_{S}^{\mathrm{max}}$ from the thermopower, and $T_{\mathrm{c}}^{\mathrm{onset}}$. Lines are guides to the eyes.
}
\label{kink}
\end{figure}

Figures~\ref{kink}(a) and \ref{kink}(b) show for $p=11.2$~GPa, a decrease in $T_{\mathrm{M}}^{\mathrm{mod}}$ with increasing magnetic field $\mathbf{H}\parallel\mathbf{c}$, revealed via $\rho$ and $d\rho/dT$.
$T_{\mathrm{M}}^{\mathrm{mod}}$ becomes more field-resistant with increasing pressure: while $T_{\mathrm{M}}^{\mathrm{mod}}$ is suppressed by roughly 9~T at $p=11.2$~GPa, the anomaly seems to persist well above 10~T at 13.6~GPa [see Fig.~\ref{kink}(c)].
Similarly, the \textit{H} dependence of the N\'{e}el temperature $T_{\mathrm{N}}$ [e.g., for $p=13$~GPa in Fig.~\ref{kink}(c)] reveals an unusual behavior of magnetism in CeAu$_2$Si$_2$.
Despite $T_{\mathrm{N}}$ rapidly decreasing with increasing pressure, the magnetic phase becomes much more stable against the magnetic field.
At zero pressure, $T_{\mathrm{N}}=10$~K and the metamagnetic transition field is $H_{\rm c}=5$~T, but at pressures above 7~GPa, where $T_{\mathrm{N}}\approx 3$~K, the AF phase survives far beyond 8~T.
A similar $p$-induced increase in $H_{\mathrm{c}}$ has been observed for CeRh$_2$Si$_2$ \cite{knafo17} and seems to be related to a $p$-induced modification of the magnetic structure \cite{kawarazaki00}.
Figure~\ref{kink}(d) shows that the field dependences of $T_{\mathrm{M}}^{\mathrm{mod}}$ deduced from resistivity and calorimetry are identical, indicating that the corresponding anomalies are caused by the same phenomenon: the transition at $T_{\mathrm{M}}^{\mathrm{mod}}$.
Additionally, the temperature $T_{S}^{\mathrm{max}}$ of the maximum thermopower shows exactly the same behavior.
The field dependence of $T_{\mathrm{M}}^{\mathrm{mod}}$ is similar to that of $T_{\mathrm{N}}$ but clearly different than that of the superconducting $T_{\mathrm{c}}^{\mathrm{onset}}$.
Following a recent study of CeIn$_3$ \cite{ebihara04}, we assume the relationship $T_{\mathrm{N}}(H)=T_{\mathrm{N,}0}[1-(H/H_{\mathrm{c}})^2]$, where $T_{\mathrm{N,}0}$ is the zero-field transition temperature and $H_{\mathrm{c}}$ is the critical field.
At $p=13$~GPa, the extrapolated $H_{\mathrm{c}}$ value for $T_{\mathrm{N}}$ is roughly 30~T and an analogous relationship for $T_{\mathrm{M}}^{\mathrm{mod}}$ yields 14 T.
The similarity of the field dependences of $T_{\mathrm{N}}$ and $T_{\mathrm{M}}^{\mathrm{mod}}$ indicates that the anomalies at $T_{\mathrm{M}}^{\mathrm{mod}}$ are induced by reconstruction of the AF order, as observed at zero pressure at $T_{\mathrm{N,3}}=8.0$~K \cite{ota09}.

\subsection{Thermoelectric power} \label{IIIb}

\begin{figure}[t]
\centering
\includegraphics[width=0.85\linewidth]{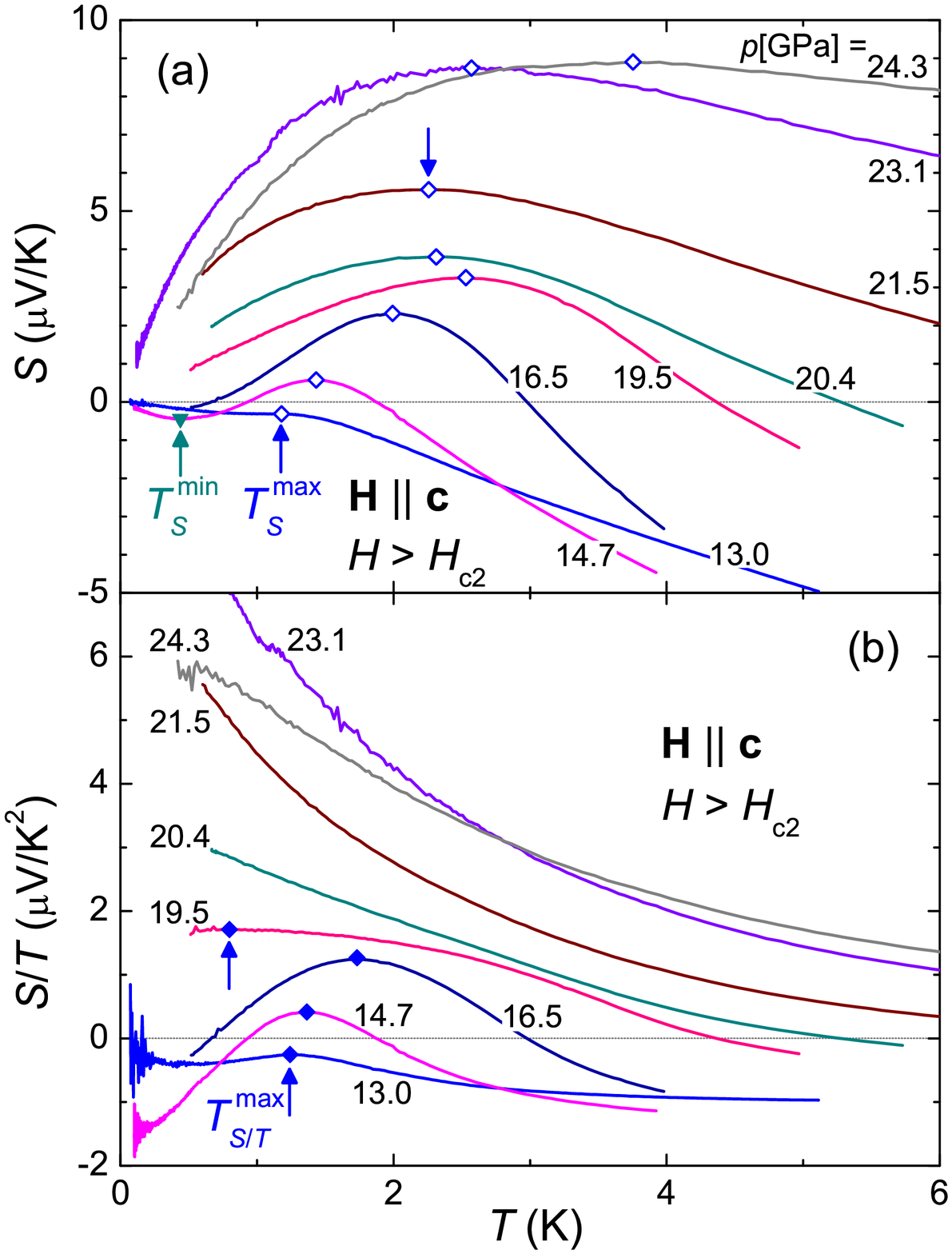}
\caption{(a) Normal-state, in-plane thermopower $S$ and (b) $S/T$ of CeAu$_2$Si$_2$ as a function of temperature $T$ at different pressures $p$ between 13.0 and 24.3~GPa, measured in magnetic fields higher than the superconducting upper critical field $H_{\mathrm{c2}}$, applied along the \textbf{c}-axis. Arrows/symbols indicate the local extrema at $T_S^{\mathrm{max}}$, $T_S^{\mathrm{min}}$, and $T_{S/T}^{\mathrm{max}}$.
}
\label{S}
\end{figure}

\begin{figure}[t]
\centering
\includegraphics[width=0.95\linewidth]{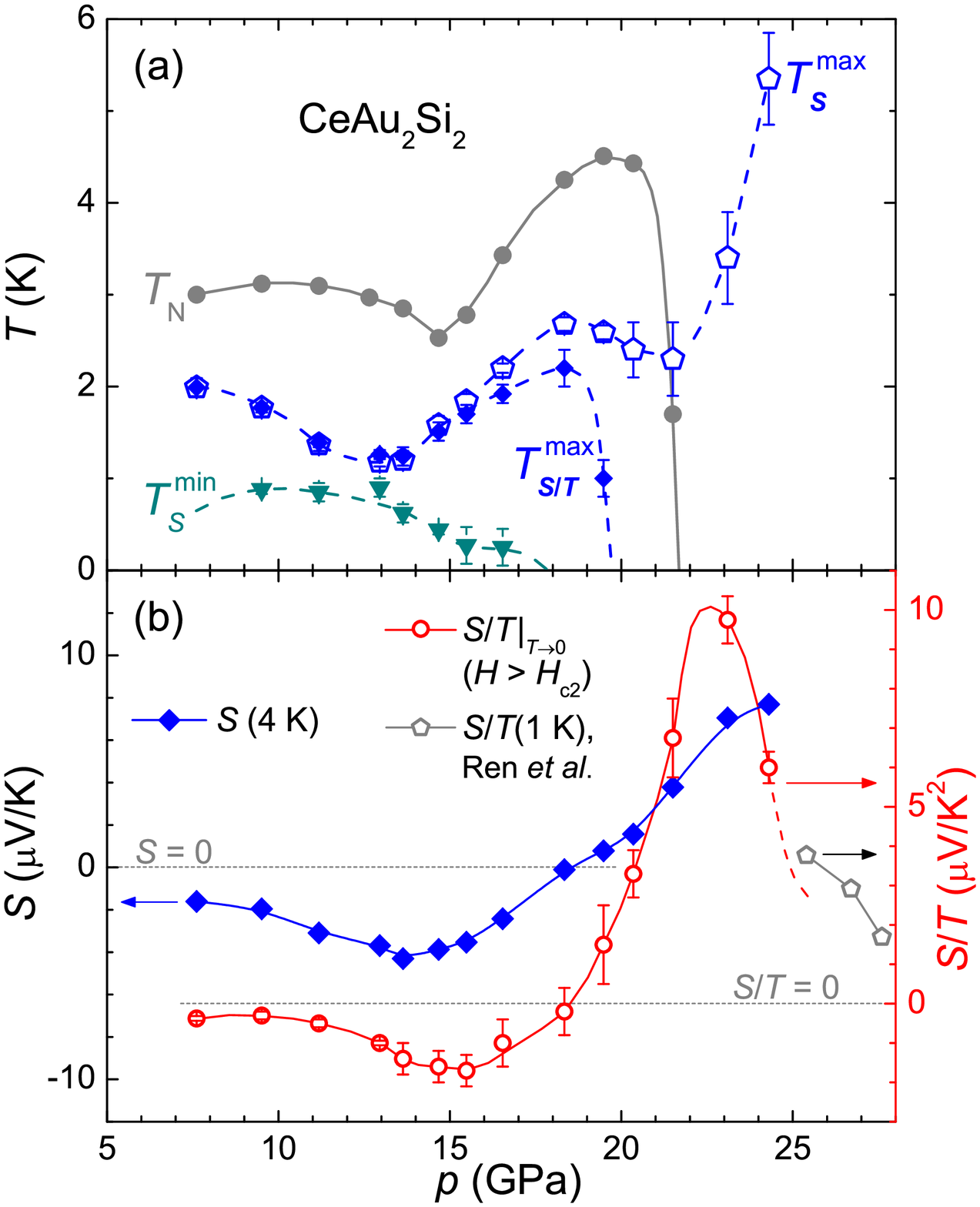}
\caption{(a) Pressure dependence of the characteristic temperatures $T_{\rm N}$, $T_{S}^{\mathrm{min}}$ (minimum $S$), $T_S^{\mathrm{max}}$ (maximum $S$), and $T_{S/T}^{\mathrm{max}}$ (maximum $S/T$).
(b) Left y-axis: thermopower $S$ at $T=4$~K. Right y-axis: $S/T$ extrapolated to zero temperature and $S/T(1$~K) from data in Ref. \cite{ren16}. Lines are guides to the eyes.
}
\label{S2}
\end{figure}

Figure~\ref{S} presents the normal-state ($H>H_{\mathrm{c2}}$) thermopower $S$ and $S/T$ as a function of temperature (below 6~K) for different pressures.
Between 7.6 and 13.6~GPa, $S$ is negative and exhibit a local maximum or a shoulder at $T_S^{\mathrm{max}}$.
Additionally, a local minimum occurs at $T_{S}^{\mathrm{min}}<T_S^{\mathrm{max}}$ for intermediate pressures [see Fig.~\ref{S2}(a) for the $p$~dependence of $T_S^{\mathrm{max}}$ and $T_{S}^{\mathrm{min}}$].
Since the local maximum becomes positive for $p\geq14.7$~GPa,$S$ crosses the zero line twice up to 16.5~GPa.
Above $T_S^{\mathrm{max}}$, $S$ decreases continuously with increasing $T$ and the temperature, at which its sign changes to negative, increases with pressure.
Furthermore, $T_S^{\mathrm{max}}$ and the magnitude of $S$ at $T_S^{\mathrm{max}}$ increase strongly with pressure when $p\gtrsim20$~GPa.
The pressure dependence of the $S$ isotherm at 4~K [Fig.~\ref{S2}(b)] is in good agreement with our previous study \cite{ren16}.
Similar to $S$, $S/T$ exhibits a maximum at the temperature $T_{S/T}^{\mathrm{max}}$, which culminates at 18.3~GPa, and vanishes abruptly at higher pressure [see Fig.~\ref{S2}(a)].
Actually, $T_S^{\mathrm{max}}\approx T_{S/T}^{\mathrm{max}}$ at low pressures, but these two temperatures diverge for high pressures.
Note that the behavior of $S$ is qualitatively different for pressures above 20~GPa, where $S/T$ decreases continuously with $T$.
For $T\rightarrow0$, $S$ tends to recover a linear \textit{T} dependence and its initial slope $S/T|_{T\rightarrow0}$ can be approximated, although systematic measurements down to still lower temperature are desirable for a better determination.
Nevertheless, Fig.~\ref{S2}(b) reveals that $S/T|_{T\rightarrow0}$ exhibits a narrow maximum of $\approx10$ $\mu$V/K$^2$ located slightly above $p_{\mathrm{c}}$.
$S/T|_{T\rightarrow0}$ continues to decrease beyond 25~GPa, as shown by data from Ref. \cite{ren16}, reflecting the increasing hybridization of Ce-4f and conduction electrons.

\begin{figure}[t]
\centering
\includegraphics[width=0.85\linewidth]{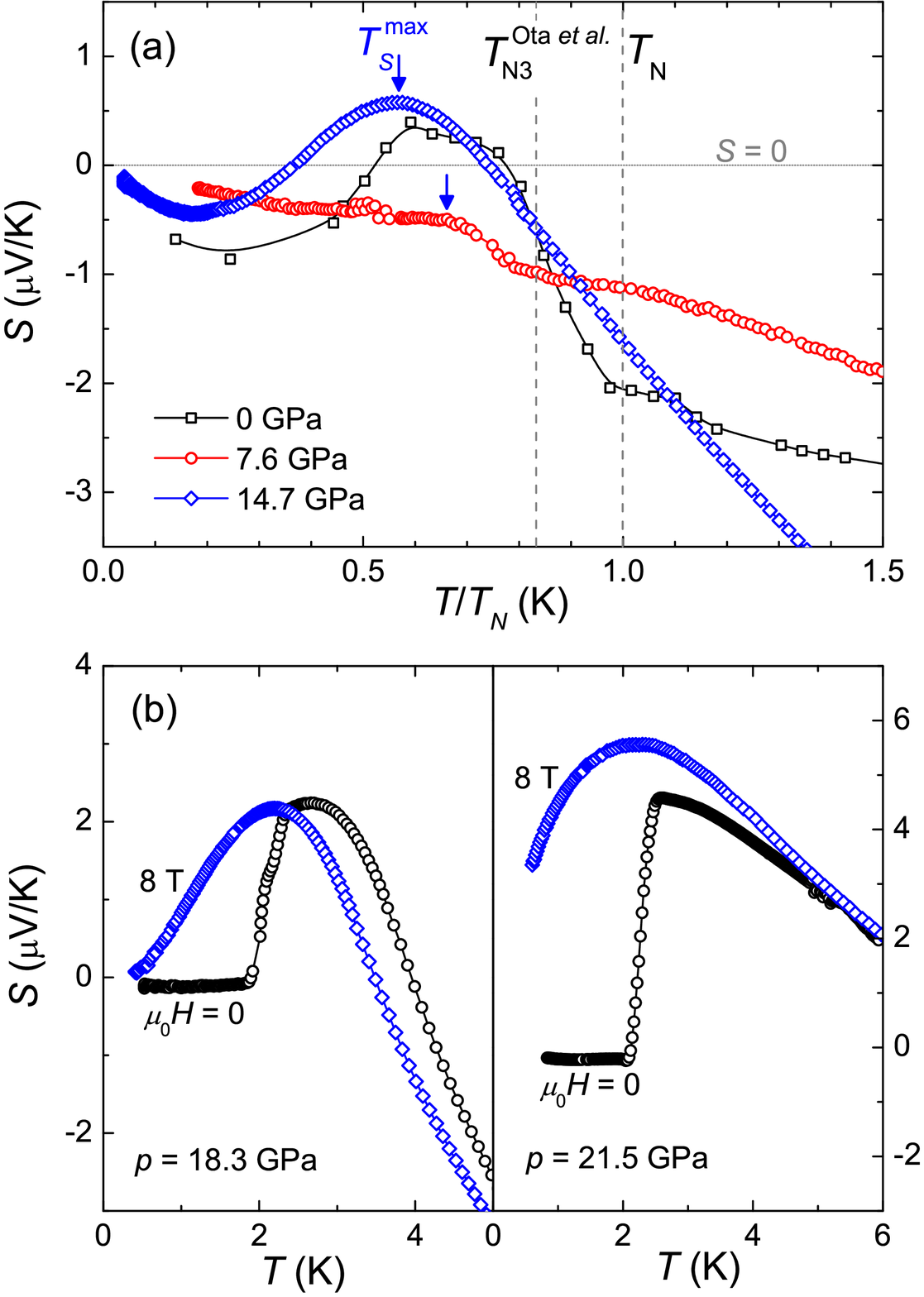}
\caption{(a) $S$ versus $T/T_{\mathrm{N}}$ for $p=0$, 7.6, and 14.7~GPa (same sample as in Ref. \cite{ren16} for $p=0$). The vertical dashed lines indicate $T_{\mathrm{N3}}$ (at $p=0$) \cite{ota09} and $T_{\mathrm{N}}$.
(b) $S$ versus $T$ at $\mu_0H=0$ and 8 T for $p=18.3$~GPa (left panel) and $p=21.5$~GPa (right panel).
}
\label{S3}
\end{figure}

Inside the AF phase, negative and positive contributions to $S$ result in subsequent local minimum and maximum, respectively, which is most clearly demonstrated for $p=14.7$~GPa.
A local extrema can be attributed to an AF gap: a minimum (maximum) is due to a gap below (above) the Fermi level and $e>0$ ($e<0$) \cite{abelskii72}.
In the present case, a putative reconstruction of the AF order with a change in the gap structure at $T_{\mathrm{M}}^{\mathrm{mod}}\lesssim T_{S}^{\mathrm{max}}$ most likely induces the oscillation of $S$ below $T_{\mathrm{N}}$.
The plot of $S$ versus $T/T_{\mathrm{N}}$ [Fig.~\ref{S3}(a)] shows that the oscillating behavior is already observed at zero pressure (data from another sample), where more marked anomalies can be brought into connection with the transitions at $T_{\mathrm{N3}}=8.0$~K and $T_{\mathrm{N}}=9.6$~K observed in the specific heat \cite{ota09}.
Thus, the features in the thermopower corroborate the interpretation of the $T_{\mathrm{M}}^{\mathrm{mod}}$ anomalies in the resistivity and calorimetry.
Furthermore, $T_{S/T}^{\mathrm{max}}$ shows field and pressure dependences [Figs.~\ref{kink}(d) and \ref{S2}(a)] very similar to those of $T_{\mathrm{N}}$ and $T_{\mathrm{M}}^{\mathrm{mod}}$, underlining their coupling.

Let us now comment on the observation that $T_{S/T}^{\mathrm{max}}$ drops to zero just above 18.3~GPa, while $T_S^{\mathrm{max}}$ persists and rises strongly at higher pressures [see Fig.~\ref{S2}(a) and also Ref. \cite{ren16}].
A feature in $S(T)$, induced by a phenomenon such as a magnetic rearrangement, is expected to persist in $S(T)/T$.
This is clearly the case of the low-pressure maximum (and minimum), whereas the high-pressure maximum occurs only in $S$ but not in $S/T$.
Therefore, the maximum in $S$ at the highest pressures is not related to a magnetic transition but rather to a crossover.
Further evidence for that the low- and high-pressure maxima have different origins is the field dependence of their magnitude [see Fig.~\ref{S3}(b)].
For $p\leq 18.3$~GPa, the magnitude of $S$ exhibits no or slightly negative field dependence.
By contrast, for higher pressures, the magnitude shows a sizable increase at $\mu_0H=8$~T.
The high-pressure maximum presumably marks a crossover between a positive zero-$T$ term and a negative intermediate-$T$ contribution.
We conclude that the low- and high-pressure maxima in $S$ have to be distinguished.

\subsection{Phase diagram and normal-state properties} \label{IIIc}

\begin{figure}[t]
\centering
\includegraphics[width=0.95\linewidth]{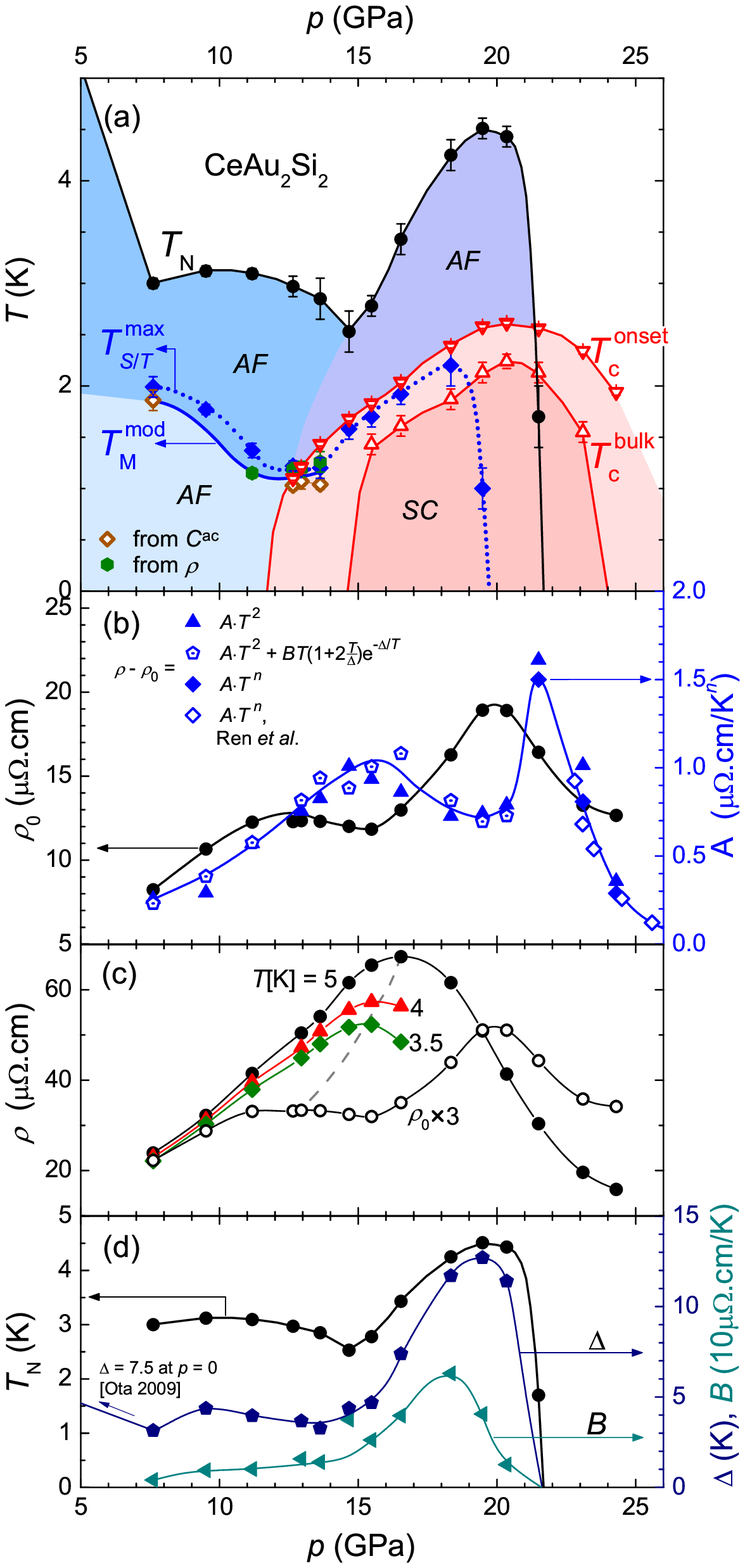}
\caption{(a) $p$-$T$ phase diagram of CeAu$_2$Si$_2$ constructed from the multiprobe data for $7<p<24.5$~GPa.
The N\'{e}el temperature $T_{\mathrm{N}}$ is extracted from $C^{\mathrm{ac}}$, $T_{\mathrm{M}}^{\mathrm{mod}}$ from $C^{\mathrm{ac}}$ and $\rho$, $T_{S/T}^{\mathrm{max}}$ from the thermopower, $T_{\mathrm{c}}^{\mathrm{onset}}$ from $\rho$, and $T_{\mathrm{c}}^{\mathrm{bulk}}$ from $C^{\mathrm{ac}}$.
(b) Pressure dependence of the residual resistivity $\rho_0$ and the power-law coefficient $A$, extracted from the normal-state resistivity ($H>H_{\mathrm{c2}}$) by fitting with $\rho(T)=\rho_0+A\cdot T^n+B\cdot T(1+2T/\Delta)\mathrm{exp}(-\Delta/T)$ ($B=0$ for $p>p_{\mathrm{c}}$). Data from Ref. \cite{ren14} are added for comparison.
(c) Isothermal $\rho$ vs $p$ at 3.5, 4, and 5~K and $3\cdot\rho_0$.
(d) Pressure dependence of the parameters $B$ and $\Delta$ compared with that of $T_{\mathrm{N}}$.
}
\label{PD}
\end{figure}

Figure~\ref{PD}(a) presents the pressure-temperature phase diagram of CeAu$_2$Si$_2$ obtained from the calorimetry, resistivity, and thermopower results.
First, let us clarify two points: i) although there is no proof from microscopic probes, we consider that the high-pressure phase below $T_{\mathrm{N}}$ is AF, which means that there is an unexpected resurgence of the AF-transition line under pressure.
This is supported by the continuity of the characteristic anomalies in the calorimetry and resistivity at $T_{\mathrm{N}}$.
ii) $T_{\mathrm{N}}$ and $T_{\mathrm{M}}^{\mathrm{mod}}$ data points below $T_{\mathrm{c}}^{\mathrm{onset}}$ are obtained from the normal state ($H>H_{\mathrm{c2}}$) and there is not yet any evidence  for the coexistence of SC and AF order.
Concerning $T_{\mathrm{N}}$ and the superconducting dome, the new phase diagram is in perfect agreement with the previous one \cite{ren14}, except for a minor downward pressure shift of $\approx1$~GPa.

The N\'{e}el temperature $T_{\mathrm{N}}$ rapidly decreases from 9.6~K at zero pressure down to 3~K at 7.6~GPa.
Here, an abrupt upturn results in a first maximum at 10~GPa. It follows a minimum ($T_{\mathrm{N}}=2.5$~K) and a second upturn at 14.7~GPa,
above which $T_{\mathrm{N}}$ strongly increases to a maximum of 4.5~K at 19.5~GPa.
Finally, the magnetic order vanishes at $p_{\mathrm{c}}\sim 22$~GPa.
Given that the pressure gradient along the sample is $\approx0.8$~GPa at $p_c$, the sudden collapse of $T_{\mathrm{N}}$ is first-order like.
The transition line $T_{\mathrm{M}}^{\mathrm{mod}}$ is established from 7.6 to 13.6~GPa. Starting at $T_{\mathrm{M}}^{\mathrm{mod}}=1.85$~K, it decreases to $\sim1.1$~K.
The behavior of $T_{\mathrm{M}}^{\mathrm{mod}}$ clearly mimics that of $T_{\mathrm{N}}$ (at about 1~K lower), indicating a magnetic origin.

The strongly modified field dependence and the peculiar pressure dependence of $T_{\rm{N}}$ suggest that the low-, intermediate-, and high-pressure magnetic ground states of CeAu$_2$Si$_2$ are different, with subsequent modifications of the propagation wave vector and Brillouin zone at $\sim8$ and $\sim15$~GPa, respectively.
Magnetic order may change from long-range localized- to itinerant-moment magnetism.
As observed at a similar unit-cell volume in CeCu$_2$Si$_2$, CeCu$_2$Ge$_2$, and CeCu$_2$(Si$_{1-x}$Ge$_x$)$_2$ alloys, the high-pressure ordering wave vector may be incommensurate \cite{stockert04,knopp89,knebel96}.
In fact, the reconstruction of magnetically ordered phases induced by different control parameters is commonly observed in Ce-HF compounds.
Examples are the effect of external pressure on CePb$_3$ \cite{morin88} and CeRh$_2$Si$_2$ \cite{kawarazaki00,knafo10} or the alloying (chemical pressure) in CeCu$_2$Si$_2$ leading to the observation of three AF phases with different propagation vectors \cite{knebel96,trovarelli97}.

\begin{figure}[t]
\centering
\includegraphics[width=0.85\linewidth]{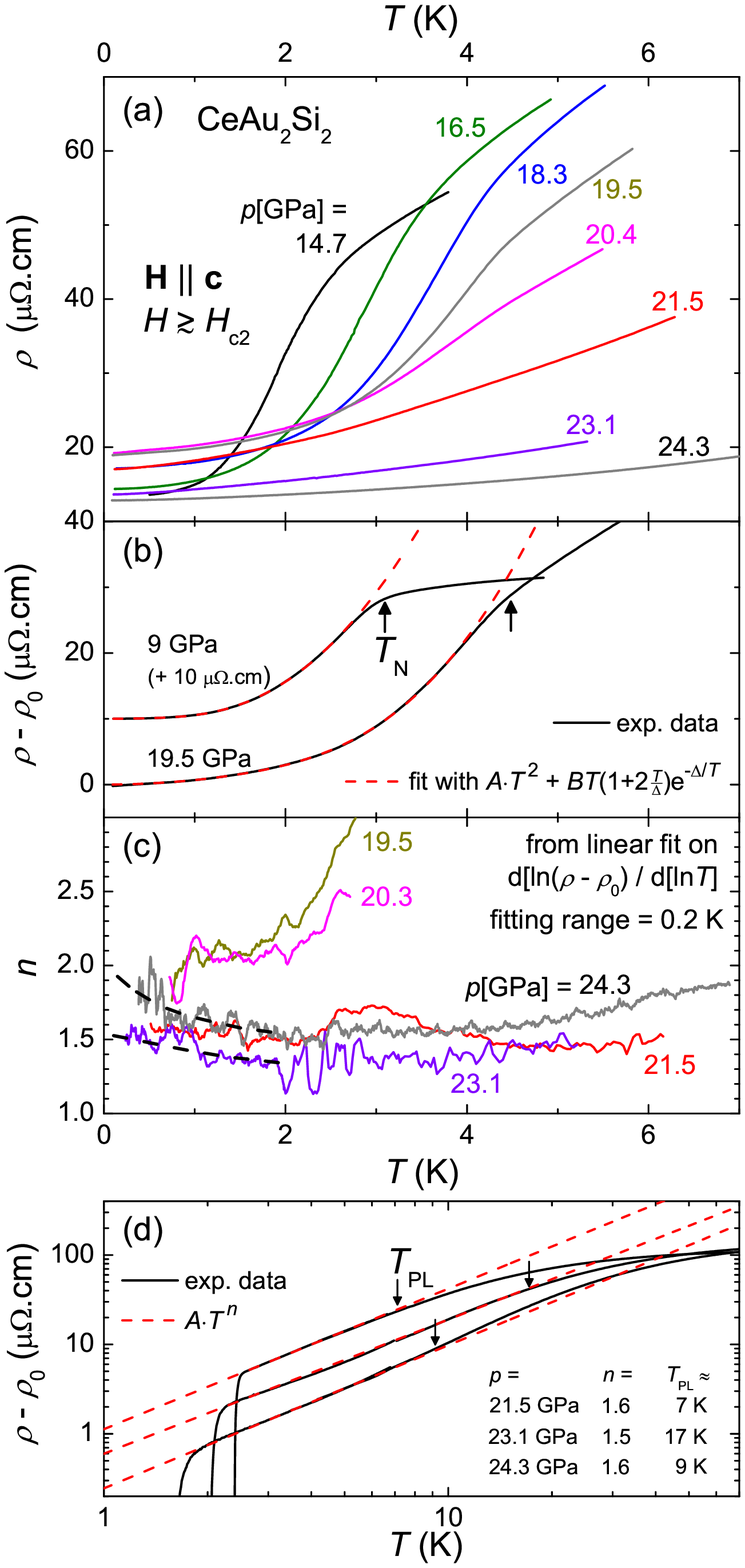}
\caption{
(a) Normal-state resistivity $\rho$ of CeAu$_2$Si$_2$ versus $T$ for $H>H_{\mathrm{c2}}$ at selected pressures $p$.
(b) Normal-state $\rho-\rho_0$ vs $T$ at 9 and 19.5~GPa. For clarity, an offset of 10~$\mu\Omega$.cm is added to $\rho$ at 9~GPa. Dashed lines represent the fitting described in the text.
(c) Temperature dependence of the power-law exponent $n$ extracted by linear fits of ln$(\rho-\rho0)$ versus ln$(T)$ over temperature ranges of 0.2~K. The dashed lines indicate extrapolations of $n$ for $T\rightarrow0$.
(d) log-log plot of normal-state $\rho-\rho_0$ vs $T$ at 21.5, 23.1, and 24.3~GPa. Dashed lines represent $A\cdot T^n$ curves.
}
\label{nln}
\end{figure}

Partial SC emerges around 12.7~GPa and $T_{\mathrm{c}}^{\mathrm{onset}}$ increases linearly almost up to the maximum of the superconducting dome.
It is most noteworthy that $T_{\mathrm{c}}^{\mathrm{onset}}$ increases simultaneously with $T_{\mathrm{N}}$ over a considerable pressure range of $\approx 6$~GPa.
Bulk SC emerges at 15.5~GPa and $T_{\mathrm{c}}^{\mathrm{bulk}}$ rises in parallel with $T_{\mathrm{c}}^{\mathrm{onset}}$ up to a maximum of 2.2~K.
The maximum $T_{\mathrm{c}}$ at 20.4~GPa is close to the collapse of magnetism at $p_{\mathrm{c}}$.
The bulk SC spans over 9~GPa and the partial SC spans a huge $p$-domain of over 17~GPa including data from Ref. \cite{ren14}.
The resistive transition is very large at the emergence and evanescence of SC.
In the region of bulk SC, the transition width is at least 0.4~K, i.e., 0.1~K more than that in Ref. \cite{ren14}, where good agreement was found between $T_{\mathrm{c}}^{\mathrm{bulk}}$ and zero-resistivity.
By contrast, the parent compounds CeCu$_2$Si$_2$ \cite{seyfarth12,giriat15} and CeCu$_2$Ge$_2$ \cite{jaccard99} exhibit narrow transitions at pressures close to the emergence ($p_{\mathrm{c}}$) or maximum ($p_{\mathrm{v}}$) of SC, while the transition is broad not only at the decrease in SC but also at intermediate pressures.
The large regions of partial SC are presumably due to exotic superconductivity, such as textured SC, as observed in CeRhIn$_5$ \cite{park12} or limited to the surface as we proposed\cite{scheerer16} for CeCu$_2$Si$_2$.

Figures~\ref{PD}(b) -- \ref{PD}(d) display salient features of the resistivity, which correlate with the two ``critical'' regions of the phase diagram, i.e., the regions of the minimum $T_{\mathrm{N}}$ and the collapse of magnetism, respectively.
As usual, the normal-state resistivity [$H>H_{\rm c2}$, Fig.~\ref{nln}(a)] is analyzed using the equation of a gapped spin-wave antiferromagnet \cite{andersen80} for $p<p_{\mathrm{c}}$ [see Fig.~\ref{nln}(b)] and the power law $\rho=\rho_0+A\cdot T^n$ for $p>p_{\mathrm{c}}$.
The residual resistivity $\rho_0$ exhibits a shoulder at 13~GPa and a maximum at 20~GPa, while the $A$ coefficient has two maxima at higher pressures by about 2~GPa [see Fig.~\ref{PD}(b)].
Below $p_{\rm c}$, due to magnetic coherent electron scattering, $A$ is reduced roughly by a factor of 8 compared with a hypothetical nonmagnetic-ground-state value, which is indeed that observed in CeCu$_2$Si$_2$ \cite{seyfarth12} and CeCu$_2$Ge$_2$ \cite{jaccard99} at their $p_{\mathrm{c}}$.
It is noteworthy that the pressure dependence of the AF energy gap $\Delta$ closely follows that of $T_{\mathrm{N}}$ [Fig.~\ref{PD}(d)].
Enhanced electronic scattering rates around 14~GPa may be related to a magnetic instability indicated by the minimum $T_{\mathrm{N}}$ and $T_{\mathrm{M}}^{\mathrm{mod}}$.
The anomaly in $\rho_0$ is relatively weak, although the shoulder in $\rho_0$ has a similar amplitude to that at $p_{\mathrm{c}}$ in the parent compound CeCu$_2$Ge$_2$ \cite{jaccard99}.
Moreover, the pressure dependence of the isothermal resistivity at temperatures above $T_{\mathrm{N}}$ [Fig.~\ref{PD}(c)], which is used to probe the ground-state excitations independently of magnetic ordering, confirms enhanced scattering around 14 GPa.
Indeed, at $T=5$~K, the $\rho$-isotherm exhibits a broad peak at 16.5~GPa, which is shifted to 15.5~GPa by lowering the temperature to 3.5~K. By continuity, this peak becomes the shoulder in $\rho_0$ at 13~GPa.
In addition, the extrapolation of the low-pressure $T_{\mathrm{N}}$ line \cite{link97,ren14} collapses at $\sim15$~GPa.
All these features including the maximum $A$ may reflect enhanced magnetic instabilities, which do not lead to a true quantum critical point (QCP) because of to the sudden strengthening of magnetism above 14.7~GPa.
Thus, only a few characteristics of spin-mediated SC are found in this $p$-region.

In the second region, i.e., around $p_{\mathrm{c}}$, the large enhancement in $\rho_0$ indicates quantum critical fluctuations \cite{miyake02}.
The maximum $\rho_0$ at 20~GPa occurs slightly below the magnetic collapse at $p_{\mathrm{c}}=22$~GPa.
The sharp anomaly in $A$ at $p_{\mathrm{c}}$ can be explained as follows. $A$ rises abruptly just below $p_{\mathrm{c}}$ since paramagnetic Ce 4f-electron scattering is recovered after the collapse of the magnetic order. 
The considerable reduction in $A$ between 21.5 and 24.3~GPa is caused by both the collapse of spin fluctuations and the increasing itinerant character of Ce 4f electrons, as in the cases of CeCu$_2$Si$_2$ and CeCu$_2$Ge$_2$ at $p_{\mathrm{v}}$ \cite{jaccard99}.
Including the data from Ref. \cite{ren14}, $A$ decreases by two orders of magnitude up to 27.6~GPa.
Furthermore, around $p_{\mathrm{c}}$, the exponent $n$ [Fig.~\ref{nln}(c)] exhibits clear NFL behavior. At $p=21.5$, 23.1, and 24.3~GPa, $n$ is close to 1.5 and constant up to roughly 10~K.
Defining $T_{\rm PL}$ as the temperature below which the power law $\rho-\rho_0=T^{n}$ with $1<n<2$ holds, it is striking that $T_{\rm PL}\approx 17$~K is maximum at $p=23.1$~GPa, i.e., slightly above $p_{\rm c}$ [see Fig.~\ref{nln}(d)].
Lastly, Fermi liquid behavior ($n=2$) is recovered below the usual temperature scale $T_{FL}$ for $p\geq24.3$~GPa.

\begin{figure}[t]
\centering
\includegraphics[width=0.95\linewidth]{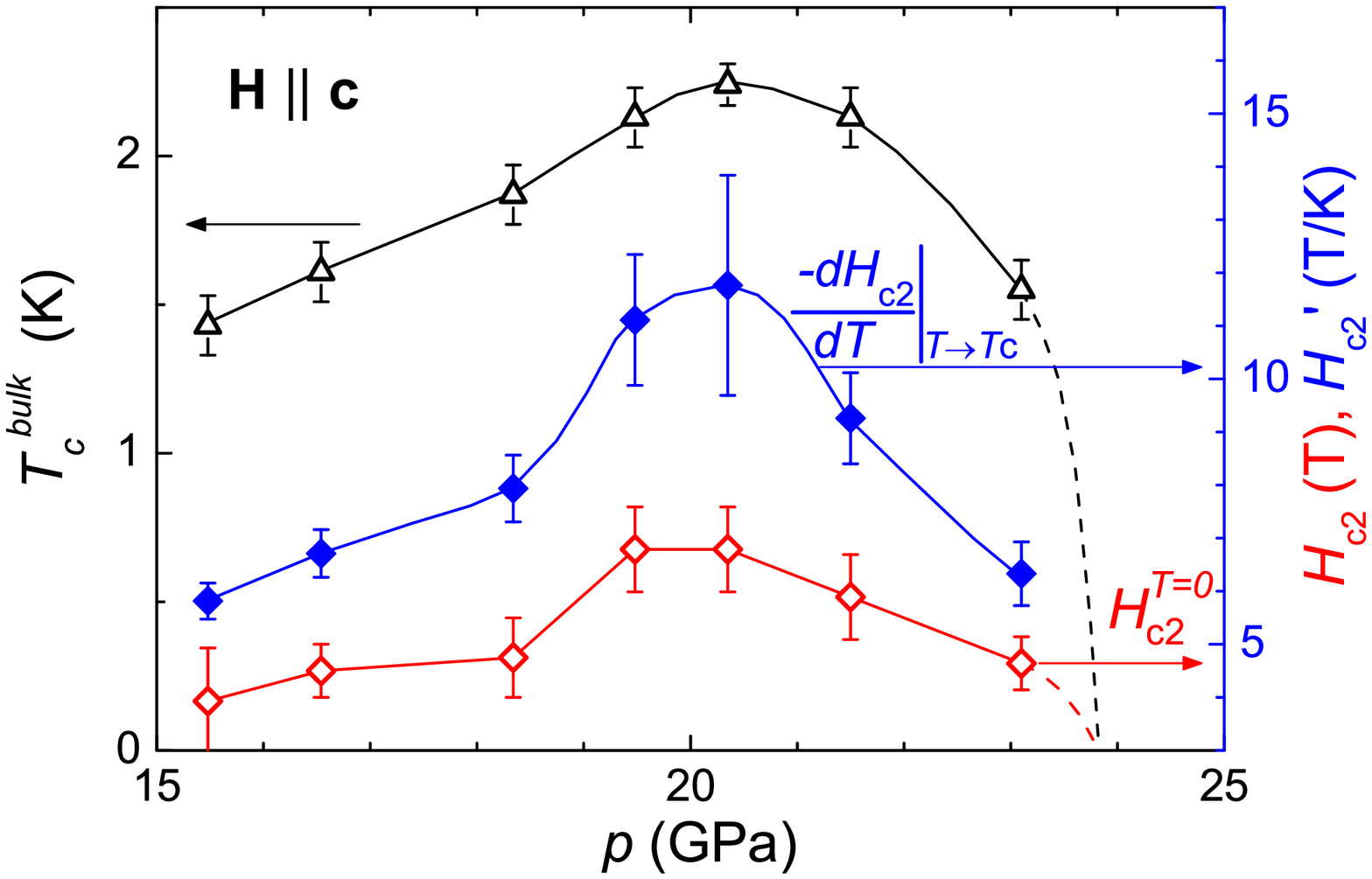}
\caption{Pressure dependence of $T_{\mathrm{c}}^{\mathrm{bulk}}$, $H_{\mathrm{c2}}(T=0)$, and $d H_{\mathrm{c2}}/d T|_{T\rightarrow T_{\mathrm{c}}}$. $H_{\mathrm{c2}}$ is defined by a 95\% drop in resistivity.
}
\label{mmm}
\end{figure}

Figure~\ref{mmm} presents properties of the upper critical field $H_{\mathrm{c2}}$ (defined by a 95\% drop in resistivity).
$H_{\mathrm{c2}}$ and its initial slope $H_{\mathrm{c2}}'=-d H_{\mathrm{c2}}/d T|_{T\rightarrow T_{\mathrm{c}}}$ show $p$ dependences similar to that of $T_{\mathrm{c}}^{\mathrm{bulk}}$, i.e., a dome shape around $p=20.4$~GPa with maximum values of 6.8 T and 11.8 T/K, respectively. From the emergence of bulk SC to the maximum $T_{\mathrm{c}}$, $H_{\mathrm{c2}}'$ is enhanced by roughly a factor of 2, indicating in --the clean limit-- an enhancement of the effective mass $m$* by 40\%.

To summarize, clear signatures of quantum critical behavior around $p_{\mathrm{c}}$ are found in the normal-state $\rho$ and in all probes sensitive to $m$*.
The quantities $A$, $S/T|_{T\rightarrow0}$, and $H_{\mathrm{c2}}'$ show similar variations, i.e., a strong peak close to $p_{\mathrm{c}}$ and a rapid decrease at higher pressures.
However, the peak positions of these quantities are spread over pressures from 20.4 to 23.1~GPa.
This difference in $p$ is rather large in comparison with the usual $p$ scale of HFs and cannot be explained alone by the pressure gradient $\Delta p<1$~GPa along the sample.
The dispersion of the peak positions over the pressure axis is significant and one can not simply attribute all the features in the normal- and superconducting-state properties to one isolated QCP. Indeed, below we discuss the interplay between the collapse of magnetism at $p_{\mathrm{c}}$ and the critical endpoint of the COV line at a slightly higher $p_{\mathrm{cr}}$.

\section{Discussion} \label{IV}

\begin{figure}
\centering
\includegraphics[width=0.9\linewidth]{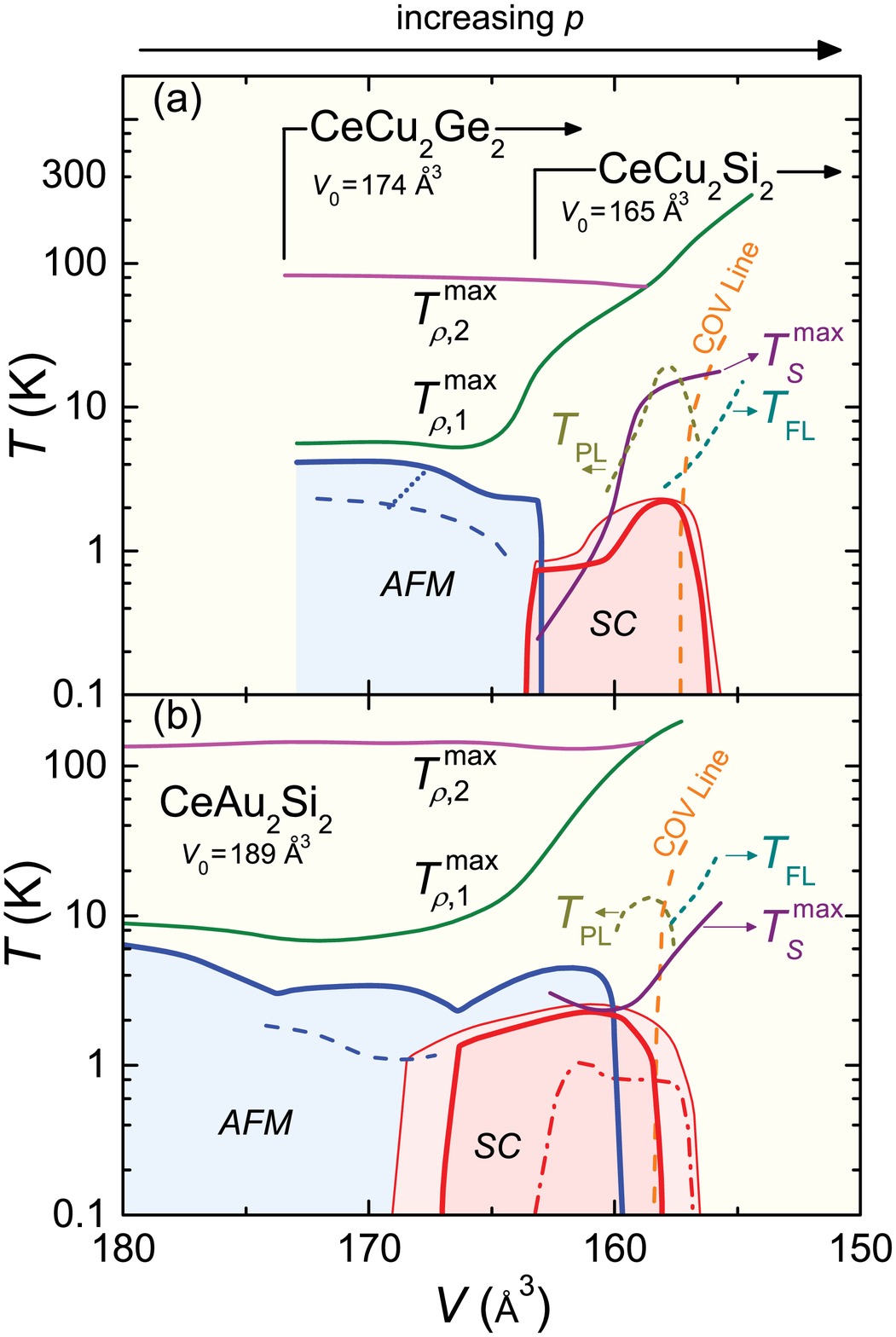}
\caption{
Schematic unit-cell-volume diagrams of (a) CeCu$_2$Ge$_2$\cite{jaccard92,link96,jaccard99,vargozthesis} ($V<173$ \r{A}$^3$) and CeCu$_2$Si$_2$\cite{jaccard85,sparn85,holmes04,seyfarth12} ($V<163$ \r{A}$^3$) and (b) CeAu$_2$Si$_2$ (from Refs. \cite{ren14,ren15}, and \cite{ren16}, and this work).
$V=189$ \r{A}$^3$ for CeAu$_2$Si$_2$ at $p=0$.
The red dashed-dotted line in (b) indicates the superconducting $T_{\mathrm{c}}^{\mathrm{onset}}$ of a self-flux-grown CeAu$_2$Si$_2$ crystal \cite{ren15}.
The scale $T_{PL}$ is defined by the temperature above which $\rho(T)$ deviates from a simple power law.
The construction of the crossover (COV) line is explained in Ref. \cite{ren15}. See text for more details.
}
\label{PDcomp}
\end{figure}

Figure~\ref{PDcomp} presents a schematic plot of the characteristic temperatures of CeCu$_2$Si$_2$, CeCu$_2$Ge$_2$, and CeAu$_2$Si$_2$ versus the unit-cell volume $V$, which highlights the striking similarities between these systems.
Such a comparison was previously elaborated in Ref. \cite{ren14}. In particular, CeCu$_2$Ge$_2$ under a pressure of $\approx 10 $~GPa coincides perfectly with CeCu$_2$Si$_2$ at ambient pressure and only data of the latter are shown for $V<163$ \r{A}$^3$.
The three compounds show a broad overlap of their superconducting regions.
The energy scales $T_{\rho,1}^{\rm max}$ due to Kondo scattering and $T_{\rho,2}^{\rm max}$ due to scattering on the excited crystal field levels (see also Fig.~\ref{intro}) merge at $V=159$ \r{A}$^3$ and the maximum $T_{\mathrm{c}}$ occurs near this volume. Thus, $T_{\mathrm{c}}$ is highest when the Kondo and CEF-splitting energies become comparable.
The Kondo temperature $T_{\mathrm{K}}$, identified as the main parameter, drives the systems from long-range magnetic ordered states, through SC, towards a strongly delocalized paramagnetic \textit{f}-metal at a reduced volume.
At pressures where the SC is strongest, the system is controlled by a relatively high energy scale: $T_{\mathrm{K}}\approx T_{\rho,2}^{\rm max}$.
Clearly, similarities in Fig.~\ref{PDcomp} underline the key role of the local Ce ion environment.
However, for a given volume $V$ the crystallographic parameters $a$ and $c$ slightly vary by $\approx 1$\% between the systems \cite{neumann85,onodera02,ohmura09}, although Cu and Au are isovalent.
Obviously Au possesses a different atomic potential from copper.
Substituting Au for Cu triples the atomic mass and increases the valence shell quantum number from 3 to 5, adding extra 5d electrons.

Only a few significant differences are found between CeAu$_2$Si$_2$ and CeCu$_2$Si$_2$/CeCu$_2$Ge$_2$.
The most important are the twofold enhancement of $T_{\rho,2}^{\rm max}$, the large overlap of SC with AF order, the magnetism spanning to a lower unit-cell volume $V$, and the SC extending to a higher $V$.
These differences require explanation and may be pertinent for understanding the nature of the superconducting pairing mechanism.
Actually, the twofold enhancement of $T_{\rho,2}^{\rm max}$ may be the crucial point, indicating that the substitution of Cu by Au significantly changes the crystal-field environment of the Ce-4f-moments, affecting the ground-state properties of the system.

In agreement with the standard behavior of Ce-based HF superconductors, the maximum $T_{\mathrm{c}}$ of CeAu$_2$Si$_2$ is close to the pressure $p_{\mathrm{c}}$ where the magnetic order vanishes.
By contrast, CeCu$_2$Si$_2$ and CeCu$_2$Ge$_2$ appear to be exceptions since their $T_{\mathrm{c}}$ peaks are far above their $p_{\mathrm{c}}$.
However, for CeCu$_2$Ge$_2$ there is one report \cite{honda13} of an ordering temperature persisting to a higher pressure, indicating a $p_{\mathrm{c}}$ closer to $T_{\mathrm{c}}^{max}$.

We now add new aspects to the comparison of CeCu$_2$(Si/Ge)$_2$ and CeAu$_2$Si$_2$.

i) The thermopower of all three systems exhibits a low-$T$ peak or shoulder. We have argued above that for CeAu$_2$Si$_2$ these features have different origins below and above $p_{\mathrm{c}}$.
For $p<p_{\mathrm{c}}$, the maximum of $S$ in CeAu$_2$Si$_2$ and CeCu$_2$Ge$_2$ \cite{jaccard92,link96} occurring below $T_{\mathrm{N}}$ is related to the opening of an AF spin gap, and its temperature scale collapses just before $p_{\mathrm{c}}$.
For $p>p_{\mathrm{c}}$, the low-$T$ maximum of $S$ occurs deep inside the coherent regime and seems to be unrelated to the Kondo effect, although its temperature scale $T_S^{\mathrm{max}}$ increases strongly with $p$.
Indeed, $T_S^{\mathrm{max}}$ in Fig.~\ref{PDcomp} (data of CeCu$_2$Si$_2$ in Refs. \cite{jaccard85} and \cite{sparn85}) is much lower than $T_{\rho,1}^{\mathrm{max}}\propto T_{\mathrm{K}}$, and is even below the Fermi liquid temperature $T_{\mathrm{FL}}$ in CeAu$_2$Si$_2$. Moreover, we note the very low $T_S^{\mathrm{max}}$ value in CeCu$_2$Si$_2$ just above $p_{\mathrm{c}}$. Concerning the origin of $T_S^{\mathrm{max}}$ for $p>p_{\mathrm{c}}$ we consider two possibilities. On one hand, $T_S^{\mathrm{max}}$ might be due to the development of magnetic correlations with decreasing $T$, or on the other hand it can signal the crossover between the zero-$T$-limit term $S\propto\gamma T$ \cite{behnia04} and the usual Kondo term as described by theory \cite{zlatic14}.

ii) Recently, Seyfarth \textit{et al}. \cite{seyfarth12} established, by resistivity scaling, the COV line in the \textit{p}-\textit{T} plane of CeCu$_2$Si$_2$ with a critical end point at $p_{\mathrm{cr}}$ and at a slightly negative temperature $T_{\mathrm{cr}}$ [see also Fig.~\ref{intro}(a)]. This crossover is ascribed to a pressure-induced change in the Ce 4\textit{f} ion valence.
Similar scaling behavior and the resulting COV line have also been reported for CeAu$_2$Si$_2$ with $p_{\mathrm{cr}}=23.6$~GPa and $T_{\mathrm{cr}}=-14$~K \cite{ren14}.
It is of upmost importance that the COV lines shown in Fig.~\ref{PDcomp} occur at almost the same $V$ for the three systems (the shift of -1 \r{A}$^3$ is within the experimental error).
This $V$ is slightly lower than that where $T_{\rho,1}^{\mathrm{max}}$ and $T_{\rho,2}^{\mathrm{max}}$ coincide.

iii) For all three systems, the most pronounced NFL behavior of $\rho(T)$ is observed in the vicinity of the COV, i.e., slightly above $p_{\rm c}$ for CeAu$_2$Si$_2$.
At $p_{\mathrm{cr}}$, the free-exponent-power-law scale $T_{\mathrm{PL}}$ has a peak value of 6 to 10 times the superconducting $T_{\mathrm{c}}^{max}$ as sketched in Fig.~\ref{PDcomp}.
For CeCu$_2$(Si/Ge)$_2$, the minimum power-law exponent is $n=1$ at the pressure of $T_{\mathrm{c}}^{max}$ \cite{jaccard99,holmes04}, while $n$ is never lower than $\approx1.5$ in CeAu$_2$Si$_2$.

iv) The red dashed-dotted line in Fig.~\ref{PDcomp} represents the superconducting $T_{\mathrm{c}}^{\mathrm{onset}}$ of a self-flux-grown CeAu$_2$Si$_2$ crystal with a similar $T_{\mathrm{N}}$ line, identical $p_{\mathrm{c}}$, but a much larger residual resistivity $\rho_0$ (compared to a low-$\rho_0$, Sn-flux-grown sample) \cite{ren15}.
The SC dome of this sample has only half the width of that of a low-$\rho_0$ sample and SC is absent for $V>163$ \r{A}$^3$, which corresponds to the observed $p_{\mathrm{c}}$ in CeCu$_2$(Si/Ge)$_2$.
Moreover, the maximum $T_{\mathrm{c}}^{\mathrm{bulk}}$ of the three systems displays the same reduction versus $\rho_0$, indicating pair breaking by nonmagnetic disorder \cite{ren15}.
Such an effect appears to be even stronger for $V>163$ \r{A}$^3$, which suggests that SC is not yet observed in CeCu$_2$Ge$_2$ for $p<10$~GPa or in partially Ge-substituted CeCu$_2$Si$_2$ \cite{knebel96,yuan03} because of the too short mean free path $l\propto \rho_0^{-1}$ in the investigated samples.

We now discuss the origin of SC in CeAu$_2$Si$_2$.
Since the first high-pressure investigation of CeCu$_2$Si$_2$, the $T_{\mathrm{c}}$ dome was first associated with the instability of the Ce ion valence \cite{bellarbi84} and subsequently ascribed to charge fluctuations \cite{jaccard99} following Miyake and coworkers, who clarified conditions for a pairing mechanism based on critical valence fluctuations \cite{miyake99,onishi00,miyake02a,holmes04,watanabe06}.
Later on, Yuan \textit{et al}. \cite{yuan03} found by Ge alloying at the Si site of CeCu$_2$Si$_2$ that the SC dome is a combination of two distinct states centered at $p_{\mathrm{c}}$ and $p_{\mathrm{v}}$.
The underlying pairing mechanism of the high-$T_{\mathrm{c}}$ region around $p_{\mathrm{v}}>>p_{\mathrm{c}}$ was thought (and this is still the dominant view) to be different from that of the low-$T_{\mathrm{c}}$ pocket around $p_{\mathrm{c}}$ \cite{ishikawa03,holmes04,holmes05,holmes07,monthoux07,lengyel09,seyfarth12}. Because of the multiple similarities (some are shown in Fig.~\ref{PDcomp}, but there are others, such as the value of the FL coefficient $A\propto1/T_{\rho,1}^{\mathrm{max}}$ \cite{ren14} and the thermopower behavior for $T<300$~K \cite{ren16}), an analogous situation is expected for CeCu$_2$Ge$_2$ and CeAu$_2$Si$_2$. By taking this approach, we first examine the emergence of SC.

One key point is that in CeAu$_2$Si$_2$, partial and bulk SC emerge at pressures close to the minimum $T_{\mathrm{N}}$.
Despite the extrapolation of $T_{\mathrm{N}}$ from $p<7$~GPa suggesting a QCP around 15~GPa, the observed magnetic line does not collapse here but at a much higher pressure.
Thus, our results diverge from the general consensus that the SC in a HF emerges near the verge of magnetism \cite{mathur98}.
The enhanced scattering rates around the minimum $T_{\rm{N}}$ [see Figs. \ref{PD}(b) and \ref{PD}(c)] support the idea that Cooper pairs are formed by magnetic fluctuations, which has been proposed to be the case for the low-$T_{\mathrm{c}}$ state in CeCu$_2$Si$_2$ \cite{stockert11,arndt11}.
However, let us recall that for this system the scenario of a spin-mediated SC, more precisely, of a nodal d-wave order parameter, has been increasingly challenged, notably by the results of thermodynamic measurements \cite{kittaka14} and theoretical calculations \cite{ikeda15}.

Then, when SC has set in, $T_{\mathrm{c}}$ considerably increases with the strengthening of magnetism over a large $V$ interval, contrary to \underline{all} previous observations.
In this broad $p$-region, it seems that SC does not compete with magnetism, as always considered but, rather, both phenomena occur in a complementary manner over an interval comparable to or larger than the total superconducting region of other HFs.
On the other hand, the partial collapse of $T_{\mathrm{M}}^{\mathrm{mod}}$ slightly precedes the emergence of SC.
Our experiment does not reveal, whether there is spatial segregation between superconducting and magnetically ordered domains or whether the SC expels magnetism.
Both cases would require that SC and magnetism do not interact on a microscopic scale.
Another possibility is that there is a dichotomy between electronic structures corresponding to different parts of the Fermi surface.

Despite the fact that the maximum of the SC dome of CeAu$_2$Si$_2$ occurs close to the vanishing of magnetism at $p_{\mathrm{c}}=22$~GPa, we do not think that the high-$p$-SC state is mainly spin-mediated.
Indeed, the following points portray a different scenario.
The unit-cell volume at which $T_{\mathrm{c}}^{\mathrm{max}}$ occurs, the $T_{\mathrm{c}}^{\mathrm{max}}$ value itself, the merging of $T_{\rho,1}^{\mathrm{max}}$ and $T_{\rho,2}^{\mathrm{max}}$ at a relatively high temperature, the proximity of the COV, and its associated NFL resistivity behavior all illustrate the leading role of valence fluctuations and their criticality at low $T$, as in the case of CeCu$_2$(Si/Ge)$_2$ \cite{jaccard99,holmes04,holmes07,miyake07,seyfarth12}.
Moreover, the collapse of $T_{\mathrm{N}}$ seems to be first-order and we think that the valence crossover COV drives the suppression of magnetism. This does not exclude the usual effect of the competition between Kondo and Ruderman–Kittel–Kasuya–Yosida interactions and the corresponding development of magnetic spin fluctuations at low $T$.
A similar scenario with a first-order collapse of magnetism has been observed for CeCu$_2$Ge$_2$ \cite{jaccard99}.
In spite of the unavoidable $p$-gradient causing $T_{\mathrm{N}}$ to spread around $p_{\mathrm{c}}$, we found a steeper drop in CeAu$_2$Si$_2$ than in CeCu$_2$(Si/Ge)$_2$ alloys, and here again lattice disorder can mitigate the collapse of $T_{\mathrm{N}}$ at $p_{\mathrm{c}}$.
Note that an analogous statement holds for the pressurized lattice CeCu$_5$Au \cite{wilhelm00} and its Ce(Cu,Au)$_6$ alloys \cite{loeneysen99}.

Critical valence fluctuations as SC-paring glue is not in conflict with the absence of traces of any valence transition in microscopic probes on CeCu$_2$Si$_2$ \cite{rueff11,kobayashi13} and CeCu$_2$Ge$_2$ \cite{kobayashi13,yamaoka14} (also because of the low resolution of these techniques).
SC is indeed predicted when the critical-end-point temperature is negative ($T_{\mathrm{cr}}<0$) \cite{miyake07,seyfarth12}, which means that only a crossover regime occurs at $T\geq0$.
A pressure gradient and sample disorder are expected to lead to an underestimation of $T_{\mathrm{cr}}$ deduced from the resistivity scaling, and one can imagine that, in a real crystal, $T_{\mathrm{cr}}$ spans over a certain range. Thus, $T_{\mathrm{cr}}$ may be positive in some islands, which signifies a trend to a local first-order valence transition with a complex nucleation process. We are now accumulating data to establish a possible relationship between $T_{\mathrm{cr}}$ and $T_{\mathrm{c}}$ for HF superconductors.

We remind readers, that two theoretical studies have independently claimed that an orbital transition and its correlated fluctuations mediate the SC in the CeCu$_2$Si$_2$ family \cite{hattori10,pourovskii14}. We have also supported this possibility for explaining the peculiar magnetic behavior of CeAu$_2$Si$_2$ \cite{ren14}, but a recent nonresonant inelastic x-ray scattering study missed a clear signature of an orbital transition in CeCu$_2$Ge$_2$ \cite{rueff15}.
Nevertheless, considering orbital physics is not unreasonable. For example, the compound PrTi$_2$Al$_{20}$, which shares some of the crucial characteristics shown in Fig.~\ref{PDcomp}, such as the merging of two resistivity contributions at the maximum $T_{\mathrm{c}}$ \cite{braithwaite}, exhibits nonmagnetic quadrupolar fluctuations \cite{matsubayashi12}.
Other exotic-order fluctuations as the possible pairing mechanism are the magnetic high-rank octopole fluctuation proposed for CeCu$_2$Si$_2$ \cite{ikeda15} and the various proposals made for the mysterious case of URu$_2$Si$_2$.

\section{Conclusion}

We have performed electric resistivity, thermoelectric power, and ac calorimetry measurements on a CeAu$_2$Si$_2$ single crystal under very high pressures and magnetic fields.
The resulting $p$-$T$ phase diagram and the normal-ground-state properties ($H>H_{\mathrm{c2}}$) exhibit new key features.
A novel magnetic-transition line $T_{\mathrm{M}}^{\mathrm{mod}}$ is associated with modifications of the AF order well below $T_{\mathrm{N}}$.
Temperature- and pressure-driven magnetic instabilities revealed by the pressure and magnetic field dependences of $T_{\mathrm{M}}^{\mathrm{mod}}$ and $T_{\mathrm{N}}$ occur in the vicinity of the SC emergence.
Strong correlations between $T_{\mathrm{N}}$, $T_{\mathrm{M}}^{\mathrm{mod}}$, and $T_{\mathrm{c}}$ are found.
The emergence of SC inside the AF phase and the maximum $T_{\mathrm{c}}$ occur in two pressure regimes, around 14 and 22~GPa, respectively, where magnetic instabilities coincide with the quantum critical behavior of the normal-ground-state properties.
In CeAu$_2$Si$_2$, $p_{\mathrm{c}}$ and $p_{\mathrm{cr}}$ are very close and it is possible that both the suppression of magnetism and the enhanced SC are driven by a valence crossover.
The COV line and the NFL behavior around $p_{\mathrm{cr}}$ indicate that the $p$-region of high $T_{\mathrm{c}}$ is governed by the proximity of a critical end point, presumably from a valence transition at negative temperature, and that the SC is probably mediated by critical valence fluctuations.
While it appears to be difficult to disentangle the phase diagram of CeAu$_2$Si$_2$, this work has made a salient progress in this direction.
Nevertheless, new theoretical work, especially CEF calculations, is required to improve the understanding of the complex magnetic and superconducting properties of CeAu$_2$Si$_2$.

\section{Acknowledgments}

We acknowledge enlightening discussions with J.-P. Brison, D. Braithwaite, G. Knebel, and J. Flouquet and technical support from M. Lopez and S. M\"{u}ller.
This work was financially supported by the Swiss National Science Foundation through Grant No. 200020-137519.

\end{document}